\documentclass[12pt,a4paper]{article}
\usepackage[pdfstartview=FitH,colorlinks=true,linkcolor=black,anchorcolor=black,citecolor=black,urlcolor=black]{hyperref}
\usepackage[english]{babel}
\usepackage{amsmath,amssymb,titling,authblk}
\usepackage{slashed}
\usepackage{amsmath}
\usepackage{amscd}
\usepackage[normalem]{ulem}
\usepackage{appendix}
\usepackage{cite}

\begin{document}

\author[1]{Oleg O.~Novikov\thanks{o.novikov@spbu.ru}}

\affil[1]{Saint Petersburg State University, 7/9 Universitetskaya nab.\\
St. Petersburg, Russia 199034}

\title{$\mathcal{PT}$-symmetric quantum field theory on the noncommutative spacetime}

\maketitle

\begin{abstract}
We consider the $\mathcal{PT}$-symmetric quantum field theory on the noncommutative spacetime with angular twist and construct its pseudo-Hermitian interpretation. We explore the differences between internal and spatial parities in the context of the angular twist and for the latter we find new $\mathcal{PT}$-symmetric interactions that are nontrivial only for the noncommutative spacetime. We reproduce the same formula for the leading order $T$-matrix of the equivalent Hermitian model as the one obtained earlier for the quantum field theory on the commutative spacetime. This formula implies that the leading order scattering amplitude preserves the symmetries of the noncommutative geometry if they are not broken in the non-Hermitian formulation.
\end{abstract}

\section{Introduction}

The unusual properties of non-Hermitian but $\mathcal{PT}$-symmetric Hamiltonians have attracted much interest recently. Because such Hamiltonians often possess real spectrum and generate an evolution that preserves a non-standard norm, they admit a pseudo-Hermitian representation and can provide a simplified description of some new class of quantum models that have a very complicated description in the usual Hermitian way \cite{BenderRealSpectra,AndrianovRealSpectra,BenderReview}. From this point of view the non-Hermitian $\mathcal{PT}$-symmetric quantum field theory (QFT) \cite{BenderSD1,BenderPhi3_1,
BenderPhi3_2,BenderCScalar,
BenderSD2,Alexandre1,Alexandre2,Alexandre3} is especially interesting because it may give a healthy interpretation for seemingly unstable and nonunitary QFTs\cite{BenderLee,BenderUnstable}. It may find many applications in the elementary particle physics \cite{AlexandreN1,AlexandreN2,OhlssonN,Braun1,Braun2} and provide many overlooked possibilities of the new physics e.g. unitary higher derivative theories of the quantum gravity\cite{MannheimCG} and stable phantom models in the cosmology
\cite{AndrianovPTom1,AndrianovPTom2,AndrianovSpringer,
NovikovPEPAN,NovikovEPJWC}.

It is interesting that for the local non-Hermitian $\mathcal{PT}$-symmetric QFT there exists a non-standard conserved inner product that preserves the Poincare invariance even though it is not in any way manifest in the commonly used construction \cite{BenderCScalar}. More so despite nonlocality of this new inner product and of the equivalent Hermitian Hamiltonian the leading scattering amplitudes of the non-Hermitian and the equivalent Hermitian models are related in an extremely simple way \cite{MyPRD}. However the nonlocality leads to causality issues that one may hope to solve by relaxing the locality of the non-Hermitian Hamiltonian \cite{JonesScatter1,ZnojilScatter1,ZnojilScatter2,ZnojilScatter3}.

To explore more how spacetime symmetries and a possible nonlocality of the non-Hermitian model are transferred to the equivalent Hermitian model, in this paper we study the properties of the $\mathcal{PT}$-symmetric QFT on the spacetime with noncommutative geometry (NCG) (for a review see Ref.\cite{NCGRev1}). There are heuristic arguments against a possibility to meausure positions with arbitrarily high precisions. According to Refs. \cite{Bronstein,Doplicher95}, such a measurement would require an extremely high concentration of energy resulting in a production of a black hole. Physical limitations on the UV behavior were considered in various contexts, e.g. in Ref. \cite{Kurkov:2013gma} the physical UV cutoff was introduced explicitly. Nevertheless, the most elegant way to incorporate the short-distance measurement limitation is given by NCG, where the uncertainty relations between the coordinates appear naturally. The NCG can be defined with the introduction of the noncommutative $\star$-product\cite{Bayen} with the $\star$-local interactions introducing certain degree of the nonlocality in the usual sense.
Among examples of such NCGs are the Gr\"{o}newold-Moyal spacetime\cite{Groenewold,Moyal,SzaboRev} that was also obtained from stringy construction\cite{stringy1,stringy2,stringy3}, the $\kappa$-Minkowski spacetime\cite{kappamink1,kappamink2} and the family of the $\mathbb{R}_\lambda^3$ spaces\cite{Rl1,Rl2,Rl3,Rl4,Rl5,RL6,RL7,RL8,RL9,RL10,RL11}. The QFT on such spacetimes is invariant under the deformations of the Poincare symmetry \cite{defsym1,defsym2} and possess unusual properties like UV/IR mixing \cite{UVIR1,UVIR2,UVIR3} and the existence of the Gribov copies in the $U(1)$ gauge theories \cite{Gribov1,Gribov2}.

However while the aforementioned simplicity and Poincare symmetry of the scattering amplitudes in the pseudo-Hermitian quantum field theory suggest that much more suitable description may exist at this point we are restricted to the Hamiltonian formulation and the operator approach. This forces certain limitations on the noncommutative geometries we may investigate. Namely we have to avoid the noncommutativity between temporal and spatial coordinates (like e.g. in the $\kappa$-Minkowski spacetime) so that one could define a slice of the constant time in a self-consistent way. In this paper we concentrate on a particular case of the noncommutative space with the angular twist that for the Hermitian QFT was extensively explored in \cite{MaxTwist}. Unlike the free QFT on the $\kappa$-deformed space, the free QFT on the space with the angular twist was shown to be equivalent to the free QFT on the ordinary Minkowski spacetime. This allows us to identify the asymptotic states with the particles on the commutative spacetime that may be important for the existence of S-matrix and may mitigate the effects of noncommutativity at large distances. Compared to the Moyal space, the $\star$-local interactions on the space with the angular twist respect a deformed version of the energy-momentum conservation law that greatly simplifies the structure of the interaction theory. These advantages make the noncommutative space with the angular twist an appropriate choice for the perturbative pseudo-Hermitian QFT construction even though it may not be the best from the phenomenological point of view.

The paper is organized as follows. In Section \ref{SecNCG} we elucidate the construction of the NCG with the angular twist by the appropriate definition of the noncommutative $\star$-product of the fields. In Section \ref{SecPT} we briefly review the basics of the pseudo-Hermitian construction for the $\mathcal{PT}$-symmetric Hamiltonians and the general scattering formula derived in \cite{MyPRD}. Then in Section \ref{SecQFT} we consider the non-Hermitian quantum field theory on the aforementioned NCG and discuss the difference between two possible parity symmetries - the internal reflection that affects only the field variables but not the coordinates and the spatial reflection. In Section \ref{SecSMatrix} we derive the formula for the leading order $T$-matrix for the $\star$-local non-Hermitian interactions. Finally in the Conclusions we give a comment on the obtained results.

\section{Noncommutative geometry with the angular twist\label{SecNCG}}

The noncommutative spacetimes may be defined by replacing the ordinary product of the fields with a deformed $\star$-product that results in the nontrivial commutation relations for the coordinates. The generic feature of this approach is that the resulting model is no longer symmetric under the ordinary Poincare symmetry which is replaced by the new deformed symmetry. Because we have to rely on the Hamiltonian approach we will restrict noncommutativity to the spatial sector. In particular we will study the noncommutative space with a Drinfel'd twist\cite{Drinfeld} that is characterized by the following commutation relations\cite{angtwist1,angtwist2},
\begin{equation}
[x^3\stackrel{\star}{,}x^1]=-i\theta x^2,\quad [x^3\stackrel{\star}{,}x^2]=i\theta x^1.\quad
[x^1\stackrel{\star}{,}x^2]=0.
\end{equation}
In the polar coordinates $x^1=\rho\cos\varphi,\, x^2=\rho\sin\varphi$
we can rewrite them as,
\begin{equation}
[x^3\stackrel{\star}{,}\rho]=0,\quad
[x^3\stackrel{\star}{,} e^{i\varphi}]=-\theta e^{i\varphi},\quad
[\rho,\stackrel{\star}{,} e^{i\varphi}]=0.
\end{equation}
To realize these commutation relations we can define the $\star$-product for the general fields as,
\begin{align}
(f\star g)(x)=\exp\left(\frac{i\theta}{2}(\partial_{y^3}\partial_{\varphi_z}-\partial_{z^3}\partial_{\varphi_y})\right)f(y)g(z)\Big\vert_{y=z=x},
\label{stardef}
\end{align}
where the derivatives are understood as the usual commuting derivatives.

The remarkable property of this particular NCG pointed out in \cite{MaxTwist} is that the $\star$-product of the plane waves can be rewritten as,
\begin{equation}
e^{-ip_1 x}\star\ldots \star e^{-ip_n x}=e^{-i(p_1+_{\star}\ldots +_{\star}p_n) x},
\end{equation}
where the $\star$-sum acts as the ordinary commutative sum for the temporal components and for the spatial components is given by,
\begin{equation}
\vec{p}_1+_{\star}\ldots+_{\star}\vec{p}_n=
\sum_{k=1}^n\hat{R}\left(\sum_{j=1}^{k-1}p_j^3-\sum_{j=k}^{n}p_j^3\right)\vec{p}_k,
\end{equation}
with the spatial twist matrix,
\begin{equation}
\hat{R}(\alpha)=\begin{pmatrix}
\cos\frac{\theta\alpha}{2}&\sin\frac{\theta\alpha}{2}&0\\
-\sin\frac{\theta\alpha}{2}&\cos\frac{\theta\alpha}{2}&0\\
0&0&1
\end{pmatrix}.
\end{equation}

Then if we use the momentum representation of the fields,
\begin{equation}
\phi_k(x)=\frac{1}{(2\pi)^{3}}\int d^3p\,\phi_{k}(\vec{p},t) e^{i\vec{p}{x}},
\end{equation}
we obtain the following simple formula for the product of $n$ fields,
\begin{align}
&\int d^3x\,\phi_1(x)\star\ldots\star\phi_n(x)\nonumber\\
&=(2\pi)^{3}
\int \left(\prod_{k=1}^n \frac{d^3p_k}{(2\pi)^{3}}
\right)\phi_{1}(\vec{p}_1,t)\ldots \phi_{n}(\vec{p}_n,t)
\delta^{(3)}\Big(\vec{p}_1+_\star\ldots +_\star \vec{p}_n\Big).
\label{MomentumRepresentation}
\end{align}
This form suggests that the quantum field theory with the $\star$-local interactions (i.e. interactions that can be written as integrals over the finite $\star$-polynomials of the fields and their derivatives in the same point of spacetime) satisfies the deformed momentum conservation law. One should also note that for the product of the two fields this formula yields that the $\star$-product trivializes,
\begin{equation}
\int d^3x\,\phi_1(x)\star\phi_2(x)=\int d^3x\,\phi_1(x)\phi_2(x).\label{FreeIsFree}
\end{equation}

\section{Scattering in the pseudo-Hermitian model\label{SecPT}}

Consider the non-Hermitian but $\mathcal{PT}$ Hamiltonian,
\begin{equation}
H\neq H^\dagger,\quad [\mathcal{PT},H]=0,
\end{equation}
where $\mathcal{P}$ is the parity operator and $\mathcal{T}$ is the time reflection operator. It happens that such Hamiltonians often have purely
real spectrum and produce the unitary evolution for the non-standard norm \cite{BenderRealSpectra}. This is related to the fact that such Hamiltonians are pseudo-Hermitian operators \cite{MostafazadehPH},
\begin{equation}
H=\eta^{-1}h\eta,\quad h=h^\dagger,\quad
H^\dagger \eta^\dagger\eta=\eta^\dagger \eta H.
\label{PHdef}
\end{equation}

We will restrict ourselves to the following class of $H$,
\begin{align}
H=\sum_{k=0}^{+\infty}(ig)^k H_k,\quad H_k=H_k^\dagger,
\quad
\mathcal{P}H_k\mathcal{P}=(-1)^k H_k,\quad
\mathcal{T}H_k\mathcal{T}=H_k,\label{Hseries}
\end{align}
where the coupling constant $g$ is assumed to be small.
For the similar decomposition of the equivalent Hermitian
operator $h$ and the intertwining operator $\eta$,
\begin{equation}
h=\sum_{k=0}^{+\infty}g^k h_k,\quad
\eta=\exp\left[-\sum_{k=0}^{+\infty}\frac{g^{2k+1}}{(2k+1)!}Q_k\right],
\end{equation}
where the following properties are assumed,
\begin{equation}
Q_k^\dagger=Q_k,\quad \{Q_k,\mathcal{P}\}=0,\quad \{Q_k,\mathcal{T}\}=0.
\label{Qextra}
\end{equation}
one obtains the following relations \cite{BenderReview},
\begin{equation}
i[H_0,Q]=H_1,\quad h_1=0,\quad h_2=-H_2-\frac{i}{2}[Q,H_1].\label{EquivH}
\end{equation}

As was pointed out in \cite{MyPRD} the first relation has a simple but powerful interpretation in the interaction picture,
\begin{align}
Q(t)=e^{iH_0t}Q e^{-iH_0t},\quad H_k(t)=e^{iH_0t}H_k e^{-iH_0t}.
\end{align}
\begin{equation}
\partial_t Q(t)= H_1(t),\quad
Q(t_2)-Q(t_1)=\int_{t_1}^{t_2}dt\, H_1(t).
\label{Qformula}
\end{equation}

This relation is in agreement with
the time-dependent pseudo-Hermiticity relation
\cite{MostafazadehNonstat,CannataNonstat,FringMoussa,ZnojilTD1,ZnojilTD2}
applied to the interaction picture,
\begin{equation}
H_I(t)=\Big[\eta(t)\Big]^{-1}h_I(t)\eta(t)-i\Big[\eta(t)\Big]^{-1}\Big(\partial_t\eta(t)\Big),
\end{equation}
\begin{equation}
h_I(t)=e^{iH_0t}(h-H_0) e^{-iH_0t},H_I(t)=e^{iH_0t}(H-H_0) e^{-iH_0t}.
\end{equation}

Consider the usual practical definition of the $S$-matrix,
\begin{align}
S_h\equiv \lim_{\substack{t_f\rightarrow+\infty\\t_0\rightarrow -\infty}}
U_h^{(I)}(t_f,t_0),\quad 
S_H\equiv \lim_{\substack{t_f\rightarrow+\infty\\t_0\rightarrow -\infty}}
U_H^{(I)}(t_f,t_0),
\end{align}
where $U_h^{(I)}$ and $U_H^{(I)}$ are the evolution operators in the interaction picture for the equivalent Hermitian and the non-Hermitian formulations correspondingly,
\begin{equation}
U_h^{(I)}(t_f,t_0)=e^{iH_0t_f}e^{-ih(t_f-t_0)}e^{-iH_0t_0},\,
U_H^{(I)}(t_f,t_0)=e^{iH_0t_f}e^{-iH(t_f-t_0)}e^{-iH_0t_0}.
\end{equation}

We will use the following decomposition in powers of the coupling constant,
\begin{align}
S_h\simeq 1+ig T_h^{(1)}+ig^2 T_h^{(2)}+\mathcal{O}(g^3),\quad
S_H\simeq 1-g T_H^{(1)}-ig^2 T_H^{(2)}+\mathcal{O}(g^3).
\end{align}
and also define the intertwining operators on the large times as,
\begin{equation}
Q_{in}=\lim_{t\rightarrow -\infty}Q(t),\quad
Q_{out}=\lim_{t\rightarrow +\infty}Q(t).
\end{equation}

Then the following relation can be obtained \cite{MyPRD},
\begin{equation}
T_h^{(1)}=0,\quad T_h^{(2)}=-\Re\Big[T_H^{(2)}\Big]-\frac{i}{2}[Q_{out},Q_{in}].
\label{Tmatrix1}
\end{equation}

\section{$\mathcal{PT}$-symmetric QFT on NCG\label{SecQFT}}

In the QFT context while $\mathcal{T}$ remains to be the usual time
reflection operator,
\begin{equation}
\mathcal{T}:\quad\phi(\vec{x})\mapsto \phi(\vec{x}),\quad
\pi(\vec{x})\mapsto-\pi(\vec{x}),\quad i\mapsto -i,
\end{equation}
the parity operator of the quantum mechanics can be generalized in two different ways. First, one can understand it as the \textit{intrinsic parity} operator $\widetilde{\mathcal{P}}$ (that reflects only fields and not spatial coordinates) acting on the canonical fields $\phi(\vec{x})$ and their momenta $\pi(\vec{x})$ in the following way,
\begin{align}
\widetilde{\mathcal{P}}:\quad&\phi(\vec{x})\mapsto -\phi(\vec{x}),\quad&
\pi(\vec{x})\mapsto-\pi(\vec{x}),\quad&i\mapsto i
\end{align}

On the other hand, one can still use the operator of the spatial reflection $\mathcal{P}$ while assuming that the fields are pseudoscalars,
\begin{align}
\mathcal{P}:\quad&\phi(\vec{x})\mapsto -\phi(-\vec{x}),\quad&
\pi(\vec{x})\mapsto-\pi(-\vec{x}),\quad&i\mapsto i
\end{align}

Many of the simple quantum field theories on the commutative spacetime are symmetric under both $\mathcal{PT}$ and $\widetilde{\mathcal{P}}\mathcal{T}$ symmetries. As a matter of fact for the $\phi^2(i\phi)^\epsilon$ model $\mathcal{P}H\mathcal{P}=\widetilde{\mathcal{P}}H\widetilde{\mathcal{P}}$ and all the important equations are the same. The difference appears for the commonly introduced operator $\mathcal{C}=\mathcal{P}\eta^\dagger\eta$ (which is a Lorentz scalar if $\widetilde{\mathcal{P}}$ but not for $\mathcal{P}$). However $\mathcal{C}$ serves an auxiliary role and contributes into all observable quantities in combination with the corresponding parity operator. However the $\widetilde{\mathcal{P}}\mathcal{T}$ symmetry may be preferable from the phenomenological point of view because $\mathcal{PT}$ is violated in the Standard Model.

Situation becomes more complicated for the quantum field theory on the NCG that is not reflection invariant. While $\widetilde{\mathcal{P}}$ obviously does not affect the twist and the order in the $\star$-product it is instructive to look in detail into the action of  other transformations on the $\star$-product of the pair of the field variables. Because $\mathcal{T}$ is antilinear and the ordinary product of the field operators is commutative,
\begin{align}
\mathcal{T}\Big(\phi_i\star\phi_j\Big)(\vec{x})\mathcal{T}
=\exp\left(-\frac{i\theta}{2}(\partial_{y^3}\partial_{\varphi_z}-\partial_{z^3}\partial_{\varphi_y})\right)\phi_i(\vec{y})\phi_j(\vec{z})\Big\vert_{\vec{y}=\vec{z}=\vec{x}}\nonumber\\
=\exp\left(\frac{i\theta}{2}(\partial_{z^3}\partial_{\varphi_y}-\partial_{y^3}\partial_{\varphi_z})\right)\phi_j(\vec{z})\phi_i(\vec{y})\Big\vert_{\vec{y}=\vec{z}=\vec{x}}=\Big(\phi_j\star\phi_i\Big)(\vec{x}).
\end{align}
Similarly the Hermitian conjugation also reverses the order of the $\star$-product,
\begin{align}
\Bigg[\Big(\phi_i\star\phi_j\Big)(\vec{x})\Bigg]^\dagger
&=\exp\left(-\frac{i\theta}{2}(\partial_{y^3}\partial_{\varphi_z}-\partial_{z^3}\partial_{\varphi_y})\right)\phi_j(\vec{z})\phi_i(\vec{y})\Big\vert_{\vec{y}
=\vec{z}=\vec{x}}
\nonumber\\&
=\Big(\phi_j\star\phi_i\Big)(\vec{x}).
\end{align}
Finally because $\vec{x}\mapsto -\vec{x}$ reverses the sign of $\partial_x$ but not of $\partial_{\phi_x}$ the spatial reflection $\mathcal{P}$ also reverses the order of the $\star$-product,
\begin{align}
\mathcal{P}\Big(\phi_i\star\phi_j\Big)(\vec{x})\mathcal{P}
=\exp\left(\frac{i\theta}{2}(\partial_{y^3}\partial_{\varphi_z}-\partial_{z^3}\partial_{\varphi_y})\right)\phi_i(-\vec{y})\phi_j(-\vec{z})\Big\vert_{\vec{y}=\vec{z}=\vec{x}}\nonumber\\
=\exp\left(-\frac{i\theta}{2}(\partial_{y^3}\partial_{\varphi_z}-\partial_{z^3}\partial_{\varphi_y})\right)\phi_i(\vec{y})\phi_j(\vec{z})\Big\vert_{\vec{y}=\vec{z}=-\vec{x}}=\Big(\phi_j\star\phi_i\Big)(-\vec{x}),
\end{align}
where in the last step we again used that the ordinary product of the field operators is commutative.

These observations are easily generalized on the product of multiple field operators. Consider the $\star$-local interaction term containing only the fields and no derivatives,
\begin{equation}
\tilde{H}^{(n)}=C_{i_1\ldots i_n}\int d^3x\,\phi_{i_1}\star\ldots\star\phi_{i_n}
\end{equation}
Then one finds the following transformation properties,
\begin{align}
&\mathcal{P}:\,C_{i_1\ldots i_n}\mapsto (-1)^nC_{i_n\ldots i_1},\quad
&\mathcal{T}:\,C_{i_1\ldots i_n}\mapsto C^\ast_{i_n\ldots i_1},\\
&\widetilde{\mathcal{P}}:\,C_{i_1\ldots i_n}\mapsto (-1)^{n}C_{i_1\ldots i_n},\quad
&(\cdot)^\dagger:\,C_{i_1\ldots i_n}\mapsto C^\ast_{i_n\ldots i_1}
\end{align}
Thus if $\tilde{H}^{(n)}$ gives contribution to $H_k$ satisfying the assumptions in Eq.~(\ref{Hseries}) depending on the choice of $\mathcal{P}$ or $\widetilde{\mathcal{P}}$ we obtain the following constraints on the coupling constants,
\begin{equation}
\mathcal{PT}:\,C_{i_1\ldots i_n}=(-1)^{n+k}C^\ast_{i_1\ldots i_n},\quad
\widetilde{\mathcal{P}}\mathcal{T}:\,C_{i_1\ldots i_n}=(-1)^{n+k}C_{i_1\ldots i_n},
\end{equation}
As result if the theory contains several fields the $\widetilde{\mathcal{P}}\mathcal{T}$ symmetry of the $\star$-local interactions as usual requires the $\tilde{\mathcal{P}}$-odd interaction to contain the odd number of fields. In contrast the $\mathcal{PT}$ symmetry admits more general interactions e.g. the following interaction is allowed in the QFT on the NCG but is trivial on the commutative Minkowski spacetime,
\begin{equation}
H_1=\int d^3x\,i\Big(\phi_1(x)\star\phi_2(x)\star\phi_3(x)\star\phi_4(x)-\phi_4(x)\star\phi_3(x)\star\phi_2(x)\star\phi_1(x)\Big).\label{Phi4PT}
\end{equation}
In the rest of the paper the difference between $\mathcal{P}$ and $\widetilde{\mathcal{P}}$ will not be important as all our equations will transfer the parity properties of $H_1$ to the parity properties of $Q$ independently of this choice.

\section{General formula for the $S$-matrix\label{SecSMatrix}}

We will assume that the free theory is given by,
\begin{align}
S_0=\int d^4x\Bigg[\frac{1}{2}\partial_\mu\phi_j\star\partial^\mu\phi_j
-\frac{m_j^2}{2}\phi_j\star\phi_j\Bigg],
\end{align}
However from the Eq.~(\ref{FreeIsFree}) it follows that this action is equivalent to the action of the free scalar QFT on the ordinary Minkowski spacetime. Therefore the free fields can be decomposed into the usual creation and annihilation operators,
\begin{align}
\phi_j(\vec{x})=\int\frac{d^3k}{(2\pi)^3} \frac{1}{\sqrt{2E^{(j)}_{\vec{k}}}}
\left(a_{\vec{k}}^{(j)\dagger} e^{-i\vec{k}\vec{x}}+
a_{\vec{k}}^{(j)} e^{i\vec{k}\vec{x}}\right),\\
\pi_j(\vec{x})=\int\frac{d^3k}{(2\pi)^3} i\sqrt{\frac{E^{(j)}_{\vec{k}}}{2}}
\left(a_{\vec{k}}^{(j)\dagger} e^{-i\vec{k}\vec{x}}-
a_{\vec{k}}^{(j)} e^{i\vec{k}\vec{x}}\right),\label{picanonical}
\end{align}
where $E^{(j)}_{\vec{k}}=\sqrt{m_j^2+\vec{k}^2}$ and the operators have the following transformation properties,
\begin{equation}
\widetilde{\mathcal{P}}a_{\vec{k}}^{(j)}\widetilde{\mathcal{P}}=-a_{\vec{k}}^{(j)},\quad,
\mathcal{P}a_{\vec{k}}^{(j)}\mathcal{P}=-a_{-\vec{k}}^{(j)},\quad
\mathcal{T}a_{\vec{k}}^{(j)}\mathcal{T}=a_{-\vec{k}}^{(j)},
\end{equation}

We will assume that the interaction term $H_1$ can be obtained as a generalization of the local interaction term of the QFT in the Minkowski spacetime by replacing the ordinary product with the $\star$-product.
We expect that just like in the ordinary QFT in the perturbation theory one should be able to represent the interaction terms as combinations of the
creation and annihilation operators. Then Eq.~(\ref{MomentumRepresentation}) implies the decomposition \cite{MyPRD},
\begin{equation}
H_1(t)=\sum_{\{j_k,\varepsilon_k\}}\int \prod_k
\frac{d^3p_k}{E^{(j_k)}_{\vec{p}_k}}
\mathcal{V}_{\{j_k,\varepsilon_k\}}(\{\vec{p}_k\})
e^{i\sum_k\varepsilon_k E^{(j_k)}_{\vec{p}_k}t}
\delta^{(3)}\Bigg(\vec{P}_{\varepsilon_k}(\{\vec{p}_k\})\Bigg),
\label{Vdecomp}
\end{equation}
where similarly to \cite{MyPRD} we define $\mathcal{V}$ as some operator valued distribution constructed
as some $c$-function of $\{j_k,\varepsilon_k\}$ and $\{\vec{p}_k\}$ multiplied
on a combination of the creation and annihilation operators in accordance
with multiindex $\{j_k,\varepsilon_k\}$ so that $(j_k,+1)$
corresponds to $a_{\vec{p}_k}^{(j_k)\dagger}$ an $(j_k,-1)$
corresponds to $a_{\vec{p}_k}^{(j_k)}$. However for the NCG in question we should replace the usual total momentum with,
\begin{equation}
\vec{P}_{\varepsilon_k}(\{\vec{p}_k\})=\varepsilon_1\vec{p}_1+_\star\ldots+_\star\varepsilon_n\vec{p}_n.
\end{equation}

Just like in \cite{MyPRD} we use Eq.~(\ref{Qformula}) to derive the following intertwining operator,
\begin{align}
Q(t)=-i\sum_{\{j_k,\varepsilon_k\}}\int \prod_k
\frac{d^3p_k}{E^{(j_k)}_{\vec{p}_k}}
\mathcal{V}_{\{j_k,\varepsilon_k\}}(\{\vec{p}_k\})
\left[
\mathrm{P.v.}\frac{e^{i\sum_k\varepsilon_k E^{(j_k)}_{\vec{p_k}}t}}
{\sum_k\varepsilon_k E^{(j_k)}_{\vec{p_k}}}\right]
\delta^{(3)}\Bigg(\vec{P}_{\varepsilon_k}(\{\vec{p}_k\})\Bigg).
\label{Qfinal}
\end{align}
where $\mathrm{P.v.}$ denotes that the pole is taken in the principal value. As usual this is the particular solution and the arbitrary integral of motion of the free QFT can be added. However as in the ordinary QFT we expect that this ansatz preserves the spacetime symmetries of the non-Hermitian model.

Using the Dyson series we get that,
\begin{align}
T_{H}^{(1)}=-2\pi
\sum_{\{\varepsilon_k\}}\int \prod_k
\frac{d^3p_k}{E_{\vec{p}_k}}
\mathcal{V}_{\{\varepsilon_k\}}(\{\vec{p}_k\})
\delta^{(4)}\Bigg(P_{\varepsilon_k}(\{\vec{p}_k\})\Bigg),
\end{align}
where the total energy is defined by,
\begin{equation}
P^0_{\varepsilon_k}(\{\vec{p}_k\})\equiv\sum_k\varepsilon_k E^{(j_k)}_{\vec{p}_k}.
\end{equation}
Then if we understand $Q$ as an operator valued distribution acting on localized wavepackets in $E$ we can use the identitity \cite{Schweber},
\begin{equation}
\mathrm{P.v.}\frac{e^{iEt}}{E}=\frac{1}{2}\frac{e^{iEt}}{E+i\epsilon}
+\frac{1}{2}\frac{e^{iEt}}{E-i\epsilon}
\xrightarrow[t\rightarrow\pm\infty]{} \pm\pi i\delta(E)
\end{equation}
to obtain that like in the ordinary QFT,
\begin{equation}
Q_{out}=-\frac{1}{2}T_H^{(1)},\quad
Q_{in}=\frac{1}{2}T_H^{(1)},\quad
[Q_{out},Q_{in}]=0,
\end{equation}
and Eq.~(\ref{Tmatrix1}) takes the same form as for the local QFT on the ordinary Minkowski spacetime \cite{MyPRD},
\begin{equation}
T_h^{(2)}=-\Re\Big[T_H^{(2)}\Big].
\label{FinalT}
\end{equation}

\section{Conclusions}

In this paper we have studied the pseudo-Hermitian construction for the quantum field theory on the NCG with the angular twist. We have studied generalizations of the operators of the spatial reflection $\mathcal{P}$ and the internal reflection $\tilde{\mathcal{P}}$. While on the ordinary commutative spacetime the polynomial of the pseudoscalars transforms in exactly the same way under both reflections we have discovered that it is no longer the case for the NCG with the angular twist. Among the $\mathcal{PT}$-symmetric interactions we find new kind of the $\mathcal{P}$-odd interactions that contain an even number of fields e.g. Eq.~(\ref{Phi4PT}) that are non-existent for the scalar QFT on the commutative spacetime.

Assuming the $\star$-local ansatz Eq.~(\ref{Vdecomp}) we obtain the same relation Eq.~(\ref{FinalT}) between the $T$-matrices of the non-Hermitian and the equivalent Hermitian models as the one obtained for the QFT on the commutative spacetime \cite{MyPRD}. This equation implies that the $S$-matrix of the equivalent Hermitian model preserves the symmetries of the initial non-Hermitian model including the symmetry of the NCG. Thus we may hope that the conjectured covariant approach to the pseudo-Hermitian $\mathcal{PT}$-symmetric QFT may also apply to this type of NCG.

This raises many interesting questions whether one may construct healthier QFT on NCG by using the pseudo-Hermitian $\mathcal{PT}$-symmetric construction. This may happen thanks to the interplay between the nonlocalities of the $\star$-product and the intertwining operator and unitarization of the higher derivative terms\cite{ghost1,ghost2,MannheimCG} that hopefully will cure the longstanding UV/IR mixing problem of the QFT on NCG.
We hope that the generalization of this approach on the geometries with space-time noncommutativity like the $\kappa$-Minkowski spacetime may also help with the causality issues of the pseudo-Hermitian models though the exact nature of the appropriate NCG is yet to be investigated.
We plan to explore these possibilities in the future work.

\section*{Acknowledgments}

The author would like to thank M.A. Kurkov for helpful discussions. The work
was supported by the RFBR project 18-02-00264 as well as by the travel grant of Saint Petersburg State University CONF2019$\_$1 id:37693026.

\providecommand{\noopsort}[1]{}\providecommand{\singleletter}[1]{#1}%

\end{document}